\documentclass[aps,twocolumn,pre]{revtex4}
\usepackage{graphicx}
\usepackage{latexsym}
\usepackage{amsmath}
\usepackage{amsfonts}
\usepackage{amssymb}
\usepackage{color}

\begin{document}
\title{Electrophoresis of a DNA Coil Near a Nanopore}
\author{Payam Rowghanian}
\email{payam.rowghanian@physics.nyu.edu}
\author{Alexander Y. Grosberg}
\affiliation{Department of Physics and Center for Soft Matter
Research, New York University, 4 Washington Place, New York, NY
10003, USA}
\date{\today}

\begin{abstract}
Motivated by DNA electrophoresis near a nanopore, we consider the flow field around an ``elongated jet'', a long thin source which injects momentum into a liquid. This solution qualitatively describes the electro-osmotic flow around a long rigid polymer, where due to electrohydrodynamic coupling, the solvent receives momentum from the electric field. Based on the qualitative behavior of the elongated jet solution, we develop a coarse-grained scheme which reproduces the known theoretical results regarding the electrophoretic behavior of a long rigid polymer and a polymer coil in a uniform field, which we then exploit to analyze the electrophoresis of a polymer coil in the non-uniform field near a nanopore. 
\end{abstract} 

\maketitle

\section{Introduction}

Due to coupling between electric and hydrodynamic fields, electrophoresis of a polymer molecule in a solvent is a complex phenomenon. Although there are only a handful of rigorously solvable models \cite{henry1931cataphoresis,cleland1991electrophoretic,evans1994colloid,muthukumar1996theory}, some quite detailed knowledge has been accumulated over the years on the electrophoretic motion of either colloidal particles or polyelectrolyte molecules, such as DNA, in a uniform electric field \cite{hoagland1999capillary,desruisseaux2001electrophoresis,PhysRevLett.76.3858,ELPS:ELPS2424}. The difficulty arises when one attempts to treat more complex geometries such as, e.g., a DNA molecule driven electrophoretically into a nanopore \cite{Kasianowicz26111996}. Due to the non-uniformity of the field \cite{wanunu2009electrostatic} and restricted conformational mobility near a membrane, such a problem cannot undergo a rigorous analytical treatment and extremely challenging for simulations. Therefore, development of a scheme which provides reliable yet manageable scaling estimates, which so far is missing, seems to be necessary. In this work, we develop such a scheme by focusing on a scaling characterization of the electro-osmotic flow; in doing so, we draw an analogy between the electrically driven charged layer of liquid around the DNA chain and a weak ``submerged jet'' \cite{landau1987fluid}. This analogy is a useful picture, which allows us to consider the electro-osmotic flow as one created by the superposition of the long-range fields of many sources of externally injected momentum, or jets, and to use its properties to develop mean-field models for more sophisticated cases such as a non-uniform field. 

We develop our scheme by first introducing an ``elongated jet,'' a long source of external momentum, and revisiting the cases of a finite rod-like DNA parallel to a uniform electric field and a DNA coil in a uniform field. These considerations reproduce the known results such as the size independent electrophoretic mobility for those cases. With this scheme at hand, we then analyze the electrophoresis of a DNA molecule in the non-uniform electric field near a nanopore and find a local mean-field relation for the electrophoretic pull exerted on each DNA segment. To motivate the connection between this analysis and the theory of DNA capture into a nanopore \cite{wanunu2009electrostatic,grosberg2010dna}, we calculate the work of the stall force on a DNA coil under electrophoretic pull brought quasistatically into the pore; this work plays the role of a quasiequilibrium energy landscape, an important ingredient of the capture theory, as it characterizes the attractive role of the electric field in drawing the DNA towards the pore. The results obtained from this consideration are then applied to the DNA capture into a nanopore in a translocation experiment in an accompanying work \cite{PayamShuraCapture}. 

\section{Uniform Electric Field}

Let us begin by briefly reviewing the phenomenological model of electrophoresis \cite{PhysRevLett.76.3858}. Assuming linearity and neglecting the relaxation effect or the perturbation of the ion distribution by the electric field (the same assumptions we will make in our consideration below), the velocity of a DNA molecule subject to a uniform field $E$ and a mechanical force $F_{\mathrm{ext}}$ is written as 
\begin{equation}
v_{\mathrm{DNA}}=\mu_F F_{\mathrm{ext}}-\mu_E E.
\end{equation}
The mechanical mobility $\mu_F$ in this relation satisfies the fluctuation-dissipation relation $\mathcal{D}=\mu_F T$, where $\mathcal{D}$ is the DNA diffusion constant and $T$ is temperature. The electric term, which includes the electrophoretic mobility $\mu_E$, can be written as $(\mu_E/Q)(QE)$ by including the DNA bare charge $Q$; although $\mu_E/Q$ can be formally viewed as the mobility of the DNA driven by an electric force $QE$, it is qualitatively different from $\mu_F$ as the electric field inevitably drags the surrounding ions and thus results in electrohydrodynamic coupling. The stall force required to hold the DNA stationary ($v_{\mathrm{DNA}}=0$), $F_{\mathrm{st}}=Q_{\mathrm{eff}} E$, characterizes the strength of the electrophoretic pull and is related to $E$ through an effective charge $Q_{\mathrm{eff}}=\mu_E/\mu_F$. In this section, we revisit the relations for DNA velocity and the effective charge by considering the flow of momentum via the electrohydrodynamic flow. 

The electrophoretic mobility $\mu_E$ is known to be independent of the DNA length \cite{PhysRevLett.76.3858,ELPS:ELPS2424,cleland1991electrophoretic,muthukumar1996theory}. This is because the balance of momenta received by the negatively charged DNA chain from the electric field and from its surrounding thin ``sleeve'' of positively charged liquid is established and maintained locally on a very small length scale. The momentum flowing away from the DNA chain is smaller than what is received by the DNA and equal to zero in the absence of a mechanical force. Intuition about the charged sleeve is mostly due to Debye theory, in which charged liquid exists in a layer of thickness the Debye screening length $r_D$, usually a few nanometers in size under relevant experimental conditions \cite{hoagland1999capillary,desruisseaux2001electrophoresis}. The mere existence of this layer upon which we base our model, however, is a generic property and therefore, not affected by the applicability conditions of the Debye theory. Therefore, while because of Manning condensation \cite{manning:924}, non-mean-field and non-linear screening, and discrete DNA charge Debye theory is not directly applicable to a dsDNA, we still symbolically call the thickness of the charged layer $r_D$ and keep in mind that the applicability of our consideration is not limited to that of Debye theory.

Much insight about the flow field around a DNA moving under the influence of an electric field $E$ and a mechanical force $F_{\mathrm{ext}}$ can be obtained by looking at electro-osmosis, a long range flow field which delivers the net momentum received from the field and the mechanical force. This flow field is most simply characterized when the DNA is stalled by a force $F_{\mathrm{st}}$ (or equivalently by sitting at the DNA frame), which is where we begin. Electro-osmotic flow forms when the momentum received from the electric field by the charged sleeve is delivered partly to the far away liquid via viscous transport (partly, because some of it is delivered to the DNA chain). Every point in the charged sleeve acts like a ``submerged jet'' \cite{landau1987fluid}, a point-like source which injects momentum into a liquid and results in a long range hydrodynamic field similar to the one formed around a simple Stokes object, but with the difference being that unlike a Stokes object, a jet is fixed in space and therefore, it readily matches the case of a stalled DNA.

\subsection{\label{rigid DNA}Long rigid DNA in a uniform field}

We address a rigid DNA by first introducing a ``submerged jet'' and considering the steady state flow field around an ``elongated jet'', a long thin source of momentum built from infinitely many submerged jets. This qualitatively describes the electro-osmotic flow field around a long rigid DNA segment, based on which, we write down the momentum conservation equations. 

\begin{figure}
 \includegraphics[width=0.85\linewidth]{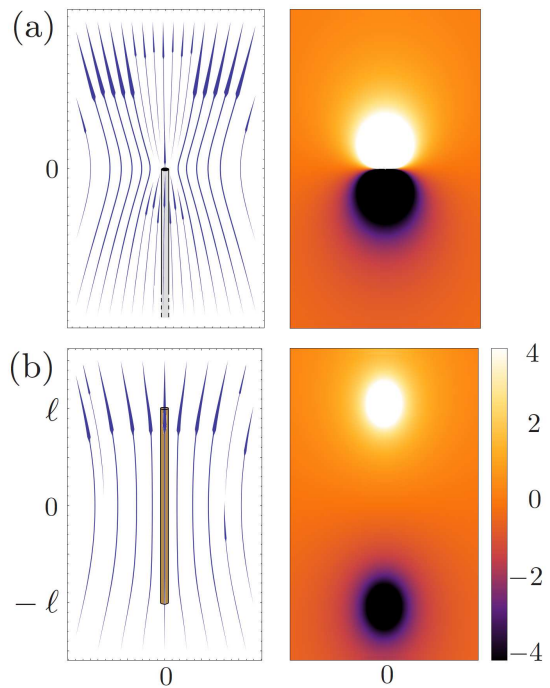}
 \caption{Velocity and pressure fields for (a) a point-like and (b) an elongated jet. Both jets drag the surrounding liquid by delivering momentum to the liquid, and create a pressure imbalance in the liquid. Pressure is positive in front of the jet and negative behind it, and diverges at the tips for infinitely thin jets in both cases. Pressure is measured in units of $\Pi/\left(8\pi \ell^2\right)$, with $\Pi$ being the momentum flux and $2\ell$ the length of the elongated jet. The overall direction of the flow is the same as the jets, and in particular, close to the elongated jet, it is almost parallel to the jet. \label{submerged single}}
\end{figure}

A submerged jet [Fig. \ref{submerged single}(a)] is created at the tip of a very long thin pipe which injects momentum at a rate $\delta\Pi$ into a liquid medium while injecting almost no liquid. The delivered momentum moves the fluid around the pipe and creates a pressure gradient between the points immediately ahead of and behind the tip of the jet. Velocity and pressure fields around a weak submerged jet ($\delta\Pi\ll \eta^2/\varrho$, where $\varrho$ and $\eta$ are liquid density and viscosity) are both linear in $\delta\Pi$:
\begin{eqnarray}
 v_r=\frac{\delta\Pi \cos{\theta}}{4\pi \eta r}&,& \ v_\theta=\frac{-\delta\Pi \sin{\theta}}{8\pi \eta r},\label{jet velocity}\\ \nonumber \\
 P&=&\frac{\delta\Pi}{4\pi r^2}\cos{\theta},\label{jet pressure}
\end{eqnarray}
where $r$ and $\theta$ are the spherical coordinates with polar axis $z$ lying along the pipe and the jet at the origin.

Let us now build an elongated jet which extends from $-\ell$ to $\ell$ along the $z$ axis, and supplies a total momentum flux $\Pi$ [Fig. \ref{submerged single}(b)]. Every small segment of length $\delta z$ provides a weak momentum flux $\delta \Pi=(\Pi/2\ell)\delta z$, where $-\ell<z<\ell$ is the position along the jet. The smallness of $\delta\Pi$ for small $\delta z$ is guaranteed by the finiteness of $\Pi$; therefore, each element $\delta z$ can be viewed as a weak point-like jet, and thus, Eqs. (\ref{jet velocity}) and (\ref{jet pressure}) represent the Green's function (also known as Stokeslet) of any extended jet, which could be integrated to obtain the pressure and velocity fields around the elongated jet:
\begin{equation}
 v_\rho=\frac{\Pi\rho}{16\pi\ell\eta}\left[\frac{1}{\sqrt{\rho^2+(z-\ell)^2}}-\frac{1}{\sqrt{\rho^2+(z+\ell)^2}}\right],\label{v rho linear jet}
\end{equation}
\begin{equation}
\begin{split}
 v_z=&\frac{\Pi}{8\pi\ell\eta}\left[\ln\left(\frac{\sqrt{\rho^2+(z-\ell)^2}+(\ell-z)}{\sqrt{\rho^2+(z+\ell)^2}-(\ell+z)}\right)\right.\\&\left. -\frac{1}{2}\left(\frac{\ell-z}{\sqrt{\rho^2+(z-\ell)^2}}+\frac{\ell+z}{\sqrt{\rho^2+(z+\ell)^2}}\right)\right],\label{v z linear jet}
\end{split}
\end{equation}
\begin{equation}
 P(z,\rho)=\frac{\Pi}{8\pi\ell}\left[\frac{1}{\sqrt{\rho^2+(z-\ell)^2}}-\frac{1}{\sqrt{\rho^2+(z+\ell)^2}}\right],\label{linear jet pressure}
\end{equation}
where $\rho$ and $z$ are the cylindrical coordinates, with the $z$ axis lying along the jet and the origin in the middle of the elongated jet. Describing the flow field everywhere around a stationary jet, Eqs (\ref{v rho linear jet}) and (\ref{v z linear jet}) coincide with the expressions for the liquid velocity at the surface of a moving slender cylindrical rod \cite{burgers1938motion,pozrikidis2011introduction}, a moving source of momentum. The hydrodynamic fields of the elongated jet derived above are linear in $\Pi$. However, this linearity, preserved formally by the smallness of the momentum flux of each small element $\delta z$, remains valid only as long as the Reynolds number for the elongated jet is small. Noting that the liquid velocity scale around the elongated jet is $v\sim \Pi/(\ell\eta)$, low Reynolds number occurs when $\Pi\ll \eta^2/\varrho$, the same as the weakness criterion for the Landau jet. 

Pressure and velocity fields for both point-like and elongated jets are shown in Figure \ref{submerged single}. Pressure is larger in front of the jets than behind them, and formally diverges at the tips of the jets. The divergence is due to zero thickness but finite momentum flux of the jets and is cut off by the finite width of a real jet. The overall motion of the liquid is in the same direction as the jet. Far from the elongated jet, at $r\gg\ell$, the elongated jet is seen as a point-like one and therefore, liquid velocity drops like $1/r$, which is the same decay as the long range flow field of a Stokes object. The resulting $1/r^2$ velocity gradient then guarantees that the momentum transfer rate is the same over any arbitrary closed surface, and thus all the jet momentum is delivered to infinity.

\begin{figure}
 \includegraphics[width=0.85\linewidth]{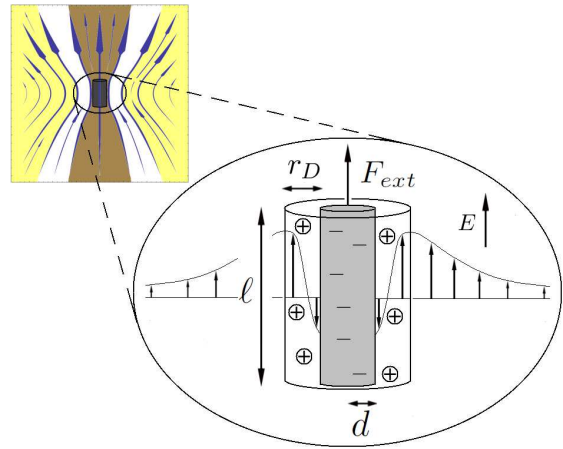}
 \caption{Flow field around a rigid DNA. A positively charged sleeve surrounding the DNA drags the outer liquid on the DNA sides through viscous shear force, and pushes (pulls) an hourglass-shaped region above (below) the DNA. Liquid velocity rapidly grows from the DNA velocity $v_{\mathrm{DNA}}$ to $v_{\mathrm{liq}}$ over a distance $r_D$, the thickness of the charged sleeve, and significantly drops over a distance of order the DNA length $\ell$. \label{cylinder}}
\end{figure}

We can qualitatively describe the elongated jet flow field as follows. First, liquid on the sides of the jet is driven by the viscous shear and its velocity drops by a significant factor at a distance about $\ell$ from the jet:
\begin{subequations}
\begin{align}
 v_z(z=0,\rho\ll\ell)&\approx\frac{\Pi}{4\pi \ell \eta}\ln{\frac{\ell}{\rho}}\label{v liq},\\
 v_z(z=0,\rho\sim\ell)&\approx\frac{\Pi}{8\pi \ell \eta}\label{v out}.
\end{align}
\end{subequations}
Second, there is a region in front of and behind the jet where liquid is driven mostly by pressure, whose hour-glass shape (Fig. \ref{cylinder}) is established from $d\rho_h/dz =v_{\rho}/v_z$ and turns out to be $\rho_h(z) \sim (d + r_D)\sqrt{z/\ell}\ll z$ for $|z| > \ell$, which is consistent with continuity and the $1/z$ drop of velocity [Eq. (\ref{v z linear jet})] along the $\rho=0$ axis [strictly speaking, at the tips, or for $0<z-\ell \ll \ell$ and $0<-(z+\ell)\ll \ell$, the form of $\rho_h$ is more complex which also results in a logarithmic correction to the relation above for $\rho_h$; we ignore this correction as it will not affect our scaling results].

Equipped with this qualitative insight, we go back to DNA electrophoresis. Let us consider a long rigid DNA of length $\ell$, radius $d$, and charge $-q$, surrounded by a thin charged sleeve of outer radius $\sim(d+r_D) \ll \ell$ which acts similar to an elongated jet, but with two complications that there is a DNA of thickness $d$ inside the jet, and that the DNA moves with some velocity $v_{\mathrm{DNA}}$. The former results in the momentum received by the sleeve being only partially delivered to infinity; in fact, the momentum delivery rate is equal to $F_{\mathrm{ext}}$, as in the absence of an external mechanical force, the total momentum received from the electric field is zero and the resulting fast decaying $1/r^3$ flow field \cite{MorrisonJr1970210,long2001note} delivers no momentum to infinity. Regarding the latter, we will justify at the end of this section the use of stationary jet for a moving DNA. 

The velocity profile, based on which we write the momentum conservation equations on different scales, is sketched in Fig. \ref{cylinder}: assuming a no-slip boundary condition on the DNA surface, liquid velocity is equal to $v_{\mathrm{DNA}}$ near the DNA, reaches $v_{\mathrm{liq}}$ over a distance of order $r_D$ from the DNA, and decays to $v_{\mathrm{out}} \ll v_{\mathrm{liq}}$ over a distance of order $\ell$. The DNA is pulled by the electric field and mechanical force $F_{\mathrm{ext}}$, and is dragged by the charged sleeve, thus
\begin{equation}
F_{\mathrm{ext}}-q E- \frac{\eta \ell}{\ln{(1+r_D/d)}}\left(v_{\mathrm{DNA}}-v_{\mathrm{liq}}\right)=0, \label{rigid-DNA force balance}
\end{equation}
where the last term is the force exerted via a viscous medium by a cylinder of radius $d+r_D$ moving with velocity $v_{\mathrm{liq}}$ on a coaxial cylinder of radius $d$ moving with velocity $v_{\mathrm{DNA}}$. The charged sleeve feels the corresponding reaction force, is pulled by the electric field, pushed from above and pulled from below by the hourglass with a force $F_{\mathrm{HG}}$, and is subject to the friction force exerted by the outside liquid; thus
\begin{equation}
\begin{split}
q E-F_{\mathrm{HG}}&-\frac{\eta \ell}{\ln{(1+r_D/d)}}\left(v_{\mathrm{liq}}-v_{\mathrm{DNA}}\right)\\&-\frac{\eta \ell}{\ln\frac{\ell}{r_D+d}}\left(v_{\mathrm{liq}}-v_{\mathrm{out}}\right)=0.
\end{split}
\end{equation}
The last term is the force exchanged between two coaxial cylinders of radii $\sim \ell$ and $r_D+d$ moving with velocities $v_{\mathrm{out}}$ and $v_{\mathrm{liq}}$ respectively.

The force $F_{\mathrm{HG}}$ exerted on the hourglass is balanced by its friction with the neighboring outside liquid. For the friction force exerted on this infinite hourglass to remain finite, it has to be dominated by the narrow parts of the hourglass near the DNA tips. Therefore, the relevant scales which determine this force are $r_D+d$, the thickness at the narrow part, and $v_{\mathrm{liq}}$, the velocity of liquid at those parts and therefore 
\begin{equation}
 F_{\mathrm{HG}}- \eta(r_D+d)v_{\mathrm{liq}}=0. 
\end{equation}
Extra logarithmic factors may be obtained by more carefully considering the shape of the hourglass and the velocity field near the DNA tips (Eqs (\ref{v rho linear jet}) and (\ref{v z linear jet})). These extra factors, however, do not affect our scaling results, as in the case of a rigid DNA, the term $F_{\mathrm{HG}}$ is small compared to others, implying that only a small fraction of momentum flows out through the hourglass and momentum is dominantly transferred via a viscous medium on the sides. This, as we will see, will not be the case anymore for the case of a bucket, for which the two contributions are comparable. 

Finally, the outside liquid is dragged by the hourglass, the charged liquid and far away walls:
\begin{equation}
-F_{\mathrm{ext}} +\eta(r_D+d)v_{\mathrm{liq}}-\frac{\eta \ell}{\ln\frac{\ell}{r_D+d}}\left(v_{\mathrm{out}}-v_{\mathrm{liq}}\right)=0. \label{rigid-out force balance}
\end{equation}
The first term in the equation above, which represents the drag force exerted by the far away walls, is crucial in satisfying Newton's third law, because the momentum flux $F_{\mathrm{ext}}$ is transferred from the DNA via the outside liquid to the walls, which exert a reaction force on the force apparatus. 

As a closely related remark, although we are building a scaling theory which is insensitive to numerical factors, every pair of terms that represent reaction forces are bound to be \emph{exactly} equal so that summing up all the equations above produces an exact triviality $0=0$, as mandated by momentum conservation. This also implies that only three out of four of the equations above are independent. We close the problem with four unknowns -- $v_{\mathrm{DNA}}$, $v_{\mathrm{liq}}$, $v_{\mathrm{out}}$ and $F_{\mathrm{HG}}$ -- with reference to Eqs. (\ref{v liq}) and (\ref{v out}), which suggest that $v_{\mathrm{out}}\sim v_{\mathrm{liq}}/\ln{[\ell/(r_D+d)]}\ll v_{\mathrm{liq}}$ and therefore, the terms containing $v_{\mathrm{out}}$ can be neglected. This yields
\begin{equation}
 v_{\mathrm{DNA}}=\frac{\ln{(\ell/d)}}{\eta\ell}F_{\mathrm{ext}}-\frac{\lambda\ln{(1+r_D/d)}}{\eta}E, \label{rigid DNA velocity}
\end{equation}
where $\lambda=q/\ell$ is the DNA charge density and all numerical factors of order unity have been dropped. The well-known mechanical mobility of a long cylinder $\mu_F\sim{\ln{(\ell/d)}}/{(\eta\ell)}$ and the electrophoretic mobility \cite{PhysRevLett.76.3858,ELPS:ELPS2424} $\mu_E\sim {\lambda\ln{(1+r_D/d)}}/{\eta}$ are reproduced in Eq (\ref{DNA velocity}). $\mu_E$ is independent of the DNA length because the bare electric force and hydrodynamic drag force both scale with $\ell$. Since $r_D$ decreases with salt concentration $c$ (as $r_D\sim c^{-1/2}$ in the Debye-H{\"u}ckel approximation), $\mu_E$ vanishes at high enough salt concentration $c$ when $r_D\ll d$, as all the momentum received by the charged sleeve is used to fully suppress the motion of the DNA, which is tightly bound by the counterions and is effectively neutral. At low salt concentrations, $r_D\gg d$ and $\mu_E\sim \ln{(r_D/d)}$, which is observed experimentally as $\mu_E \sim \ln c$ \cite{desruisseaux2001electrophoresis}.

\subsection{\label{bucket model}DNA coil in a uniform field}

\begin{figure}
 \includegraphics[width=0.99\linewidth]{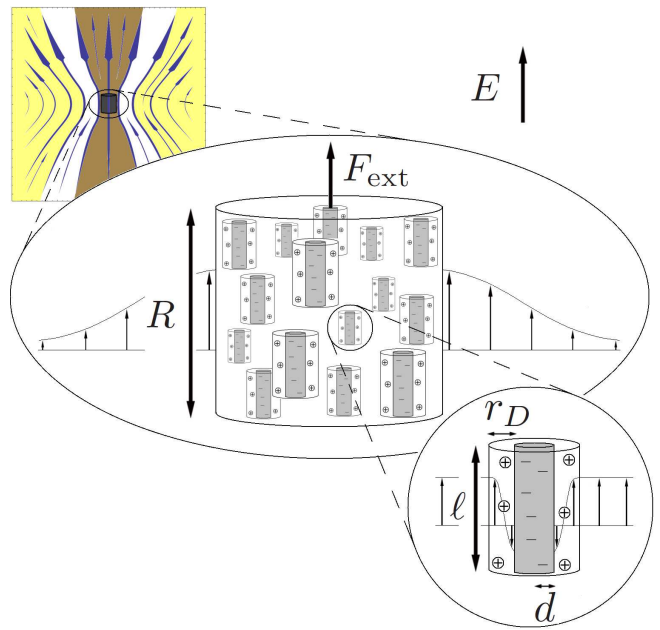}
 \caption{Top left: flow field around a bucket of DNA segments placed at the center. The liquid inside the bucket is driven by the electric field, dragging the liquid on the sides through shear friction, and pushing the hourglass through the excess of pressure on the top and bottom of the bucket. Center and bottom right: liquid velocity profile inside and outside the bucket. $N$ DNA segments of length $\ell$ and charge $-q$, surrounded by a charged sleeve of liquid of thickness $r_D$ each, are placed inside a bucket of size $R$. Inside the bucket, liquid moves with a velocity $v_{\mathrm{liq}}$ except at a distance $r_D$ from each segment, where it changes to the DNA velocity $v_{\mathrm{DNA}}$. Outside the bucket, the velocity significantly decays over a distance of order $R$.   \label{bucket}}
\end{figure}

Consider now a simple model of a DNA coil, in which $N$ long rigid DNA segments of length $\ell$ (corresponding to the DNA Kuhn length) and charge $-q$, each surrounded by non-overlapping charged sleeves of thickness $r_D$, are distributed uniformly and parallel to each other in a fictitious bucket of size $R\gg \ell$. Upon the application of an electric field, in the DNA frame, $N$ driven sleeves move and create a long-range electro-osmotic flow. Just like the case of Zimm dynamics, where $N$ mechanically pulled segments of a coil drag the liquid within the coil, the $N$ driven sleeves also collectively drag the liquid inside the bucket with a velocity $v_{\mathrm{liq}}$ (it is known that the dominant contribution to the flow inside the bucket comes from the segments around the surface and thus, small scale fluctuations of the segment density in a coil do not affect this picture). Therefore, liquid drains freely \cite{ELPS:ELPS200800257,hickey2012simulations} through the bucket and its velocity decays only outside the bucket, where it reaches a value $v_{\mathrm{out}}$ over a distance of order $R$, and at distances $\sim r_D$ and closer to each segment, where it decays to the DNA velocity (assuming a no-slip boundary condition at the DNA surface). As in the case of a rigid DNA, on the sides of the bucket, liquid is dragged by shear friction, and an hourglass above and below the bucket (Fig. \ref{bucket}) is pushed by the liquid driven inside the bucket.

Following the same approach as the one used in Sec. \ref{rigid DNA}, below we write down the analogues of Eqs. (\ref{rigid-DNA force balance}) -- (\ref{rigid-out force balance}) for a bucket. For the DNA we have
\begin{equation}
F_{\mathrm{ext}}-NqE-N\frac{\eta\ell}{\ln{\left(1+ \frac{r_D}{d}\right)}} \left(v_{\mathrm{DNA}} -v_{\mathrm{liq}}\right)=0, \label{DNA force balance eq}
\end{equation}
where in the last term, the friction force exerted on $N$ DNA segments by the corresponding sleeves has been considered. This term is indeed crucial in getting a length independent electrophoretic mobility, as it manifests the local balance of the momenta received from the electric field. For the liquid inside the bucket, we have
\begin{equation}
\begin{split}
Nq E-F_{\mathrm{HG}}-N\frac{\eta\ell}{\ln{\left(1+ \frac{r_D}{d}\right)}}\left(v_{\mathrm{liq}}-v_{\mathrm{DNA}}\right)&\\ -\eta R\left(v_{\mathrm{liq}}-v_{\mathrm{out}}\right)&=0. \label{liquid force balance}
\end{split}
\end{equation}
In the last term, friction force exerted on a body of size $R$ is found from a velocity gradient $\sim \left(v_{\mathrm{liq}}-v_{\mathrm{out}}\right)/R$ and contact area $\sim R^2$. The force $F_{\mathrm{HG}}$ exerted by the hourglass, just like the case of a rigid DNA, is dominated by the narrow parts of the hourglass and thus
\begin{equation}
 F_{\mathrm{{HG}}}- \eta R\ v_{\mathrm{liq}}=0. 
\end{equation}
Finally, for the liquid outside the bucket we have
\begin{equation}
-F_{\mathrm{ext}} +\eta R\ v_{\mathrm{liq}}-\eta R\left(v_{\mathrm{out}}-v_{\mathrm{liq}}\right)=0, \label{out force balance}
\end{equation}
where the first term represents the drag force exerted by the far away walls. To conserve momentum, as pointed out before, liquid velocity decays as $v(r)\sim v_{\mathrm{liq}}\ R/r$ and thus, $v_{\mathrm{out}}$ is smaller than $v_{\mathrm{liq}}$ by a numerical factor. Therefore, from the equations above we obtain
\begin{equation}
 v_{\mathrm{DNA}}=\frac{1}{\eta R}F_{\mathrm{ext}}-\frac{\lambda}{\eta}\ln{\left(1+ \frac{r_D}{d}\right)}E,\label{DNA velocity}
\end{equation}
where all numerical factors of order unity and a term containing $R/(N\ell)\ll 1$ have been dropped. The mechanical mobility $\mu_F=1/(\eta R)$ and the electrophoretic mobility 
\begin{equation}
\mu_E\sim \frac{\lambda}{\eta} \ln{\left(1+ \frac{r_D}{d}\right)}
\end{equation}
are reproduced in Eq (\ref{DNA velocity}); $\mu_E$ agrees with experiments \cite{hoagland1999capillary,desruisseaux2001electrophoresis}, in which, for low salt concentrations $c$ at which $r_D\gg d$, a $\mu_E \sim \ln{c}$ is observed. The ostensibly equal electrophoretic mobilities found for a DNA coil [from Eq. (\ref{DNA velocity})) and a rigid DNA [or $N=1$, Eq (\ref{rigid DNA velocity})] contain unknown numerical factors which cannot be captured using our scaling approach. Recent simulations \cite{hickey2012simulations} have produced the experimentally observed \cite{hoagland1999capillary} non-monotonic behavior of $\mu_E$ as a function of the polymerization degree $\mathcal{M}$ ($=n_{\mathrm{Kuhn}}N$, where $n_{\mathrm{Kuhn}}$ is the number of monomers in a Kuhn segment), which increases with $\mathcal{M}$ when the coil is smaller than the electric screening radius due to the overlap of ion clouds of different monomers, and then slightly decreases due to Manning condensation as $\mathcal{M}$ further increases and reaches its asymptotic value at $\mathcal{M}\approx 50$ (Figure 1 in Ref. \cite{hickey2012simulations}). This non-monotonic behavior is not captured in our consideration, because on the one hand, we have not included the change in $\lambda$ caused by the coil size dependent ion condensation and, on the other hand, even for a single rigid dsDNA segment considered in Sec. \ref{rigid DNA}, the polymerization degree $\mathcal{M}$ is already well above $50$. 

Letting $v_{\mathrm{DNA}}=0$, the stall force is found to be
\begin{equation}
 F_{\mathrm{st}}=\left(\lambda \ln{\left(1+\frac{r_D}{d}\right)} R\right) E, \label{bucket stall force}
\end{equation}
where the effective charge $Q_{\mathrm{eff}}=\lambda \ln{\left(1+{r_D}/{d}\right)} R$ scales with the bucket size and is much smaller than the bare charge $Nq$ for $N\gg 1$, because of the local balance of the momentum received from the electric field. In fact, only a small fraction $\sim \ln{\left(1+{r_D}/{d}\right)} R/(N\ell)$ of the total momentum $NqE$ received by the coil from the field flows out of the surface of the bucket to the outside liquid via shear friction and through the hourglass. The system is in this sense analogous to an electric circuit in which the battery has an internal resistance much larger than the load and thus, energy dissipation and ``voltage drop'' (balance of momentum) mostly occurs inside the battery rather than inside the load (liquid flowing outside the bucket). 

We have addressed the case in which the DNA segments are parallel to $E$ so far. Similar consideration for a DNA segment perpendicular to $E$ is possible (details not shown) by building a ``transverse jet'', a line of point-like jets sitting side by side. Conceivably, arbitrary orientation of the DNA between these two cases will only change the mobility by a numerical factor, as shown in previous works \cite{abramson1942electrophoresis,schellman1977electrical}. This then allows us to assume a more realistic model of a DNA coil as well, in which DNA segments take arbitrary orientations.

We end this section by recalling that we have approximated the flow field around a DNA by the steady state flow field of some jets. For this to be justified, the steady state flow field of the jet must have enough time to build up around the DNA as it moves. Since the liquid velocity drops significantly over a distance of order DNA size ($\ell$ for a rigid DNA and $R$ for a coil), DNA dynamics is only sensitive to the flow field at distances smaller than the DNA size. Therefore, the steady state approximation is valid if the time $t_h$ that it takes hydrodynamic perturbations to propagate to a distance comparable to the DNA size is shorter than the time $t_D$ it takes the DNA to move a distance of its own size. 

The time $t_h$ is determined from dimensional analysis to be $t_h\sim \ell^2 \varrho/\eta$ for a rigid DNA and $t_h\sim R^2 \varrho/\eta$ for a coil, where $\varrho$ is the liquid density. Using Eqs (\ref{rigid DNA velocity}) and (\ref{DNA velocity}), we find $t_D\sim \ell \eta/(E\lambda)$ for a rigid DNA and $t_D\sim R \eta/(E\lambda)$ for a coil (assuming $\ln{(1+r_D/d)}\sim 1$). The condition $t_h<t_D$ then yields $\lambda \ell\ E\varrho<\eta^2$ and $\lambda R\ E\varrho<\eta^2$. Using $\lambda\sim 1 \mathrm{e}/\mathrm{nm}$, $E\sim 10^3 \mathrm{V}/\mathrm{m}$, $\ell=100\mathrm{nm}$, $R\sim 10^4 \mathrm{nm}$ and considering water as a solvent, we get  $\lambda \ell\ E\varrho/\eta^2\sim 10^{-5}$, and $\lambda R\ E\varrho/\eta^2\sim 10^{-3}$, both strongly satisfying the steady state criterion. 

\section{\label{electrophoresis near pore}Non-uniform Electric Field Near a Pore}

\subsection{Derivation of the stall force}

Let us now consider a DNA coil near the membrane with one end held inside a pore on a membrane. The electro-osmotic flow is driven by a non-uniform electric field $E(r)=Q_{\mathrm{pore}}/r^2$ \cite{wanunu2009electrostatic}, where the pore effective charge $Q_{\mathrm{pore}}$ depends on the voltage $\Delta V$ across the apparatus and the width $a$ and depth $b$ of the pore and is $Q_{\mathrm{pore}}\simeq\Delta V a^2/(8 b)$ for $b\gg a$. In the case of a DNA coil stalled in a uniform field, the electro-osmotic flow depended linearly on the electric field, which, using Eqs. (\ref{out force balance}) and (\ref{DNA velocity}), could be found to be
\begin{equation}
 v_{\mathrm{liq}}^{\mathrm{st}}=\mu_E E. \label{liquid velocity}
\end{equation}
Equation (\ref{liquid velocity}) suggests that in a non-uniform electric field, liquid is driven at different velocities in different places; a local application of this equation, however, does not necessarily produce the flow field correctly because the liquid flow must also obey continuity.

It is an interesting coincidence that both the electric field (which is proportional to the electric current density \cite{wanunu2009electrostatic}) and the velocity field in the case of a DNA coil with one end held in the pore are dictated by the spherical geometry of the apparatus, and both decrease like $1/r^2$ with the distance $r$ from the pore. Because of such a coincidence, in this particular case, Eq. (\ref{liquid velocity}) can be readily applied as a local relation to determine the electro-osmotic flow field $v_{\mathrm{liq}}(r)$ as a function of $E(r)$. Directly from this local characterization of the electro-osmotic flow follows also a local application of Eq. (\ref{bucket stall force}): conformation of a DNA with one end in the pore is similar to that of a polymer grafted from one end to a surface \cite{vanderzande1998lattice,grimmett2001probability}, for which, the local blob size is proportional to the distance $r$ from the pore. The stall force exerted on a blob of size $r$ is proportional to its size [Eq. (\ref{bucket stall force})] and therefore, the average stall force per DNA segment at a distance $r$ can be found to be
\begin{equation}
 f_{\mathrm{st}}(r) = q_{\mathrm{eff}}(r) E(r) \sim \mu_E \eta\ell \left(\frac{r}{\ell}\right)^\frac{\nu-1}{\nu} E(r),\label{stall force per segment}
\end{equation}
in which, we have used the fact that a blob of size $r$ contains $\sim (r/\ell)^{1/\nu}$ DNA segments, with $\nu$ the Flory exponent of the coil. The effective charge per segment is $q_{\mathrm{eff}}(r)=\mu_E/\mu_F(r)$, where $\mu_E$ is the size independent electrophoretic mobility, and $\mu_F(r)$ can be formally viewed as the mechanical mobility of a single segment at $r$, inversely proportional to the friction coefficient per segment $\xi(r)\sim \eta \ell (r/\ell)^{(\nu-1)/\nu}$ in a blob of size $r$. The relation above states that the segment effective charge is larger closer to the pore, which means that the screening effect is suppressed more effectively by the membrane closer to the pore.

\begin{figure}
 \includegraphics[width=0.99\linewidth]{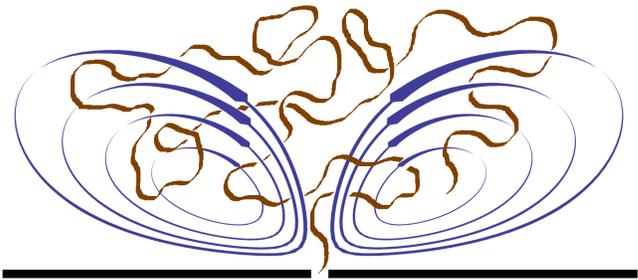}
 \caption{Electro-osmotic flow around a DNA coil held at the non-uniform electric field near the pore.  Due to the large hydraulic resistance of the pore, for the liquid to flow through the coil, it must be sucked from near the membrane and therefore, the field lines are closed as liquid circles back around the DNA by sliding beside the membrane. Constrained by mass conservation, liquid velocity drops as $1/r^2$, showing the same dependence as the electric field on the distance from the pore $r$. This allows us to obtain a local relation for the electrophoretic pull per DNA segment as a function of $r$. \label{near membrane}}
\end{figure}

Below we describe the electro-osmotic flow through the coil. Placing the DNA end inside the pore not only subjects the coil to a non-uniform field, but also brings it near a membrane which constrains the electro-osmotic flow around the DNA. For a very short pore (drilled on a thin membrane such as a graphene membrane), as we show in Sec. \ref{pore details}, the liquid driven through the coil can be partly sucked from the other side of the membrane through the pore. For a long pore, however, friction of the liquid with the pore walls is very large and thus, electro-osmotic flow can not be maintained by sucking the liquid through the pore. Instead, the flow lines are dominantly closed only on one side of the membrane, where liquid circles around the DNA by sliding along the membrane (see the sketch in Fig. \ref{near membrane}). Two comments regarding the circulation of the electro-osmotic flow are in order.

First, the liquid circulation is well facilitated by the conformation of the DNA with one end held in the pore (which we show in Sec. \ref{deformation} to be not perturbed by the electric field in the field range used in experiments). Any blob of size $r$ of a self-similar coil contains an empty region (or a void) of size $\sim r$, and thus, the electro-osmotically driven flow through those parts of the blob occupied by the DNA segments can circulate back through the voids. The circulation, of course, does not occur in the highly dissipative form of many small loops; instead, the liquid flows in one large loop away from the pore through the regions occupied by the DNA and flows back towards the pore through the empty parts. It is known that on average, the segment density of a grafted polymer near the membrane is much lower than near the axis normal to the membrane at the grafting point (or the axis of the pore in our case), and in this sense, on average, the liquid flows through the coil close to the pore axis and circles back beside the membrane, as shown schematically in Fig. \ref{near membrane}. Both the forward and backward branches of the loop pass through cross-sections whose areas scale as $r^2$, and thus, Eq. (\ref{stall force per segment}) remains valid. 

Second, the effect of the circulation on the stall force is negligible. The DNA conformation is similar to a coil placed inside a cone with an opening angle equal to $\pi$ (a special case of a polymer attached to the tip of a cone \cite{maghrebi2011entropic}), in which, any blob of size $\sim r$ is confined by the membrane to a region of size $\sim r$. Away from a confining membrane, momentum would flow out of the blob at a rate $\sim \eta r v_{\mathrm{liq}}(r)$, which will change only by numerical factors when the coil is surrounded by a surface at a distance $\sim r$ from the coil. The local momentum balance rate $\sim \eta\ell n(r)  v_{\mathrm{liq}}(r)$ [where $n(r)$ is the number of segments in the blob], therefore, will continue to be the dominant factor in determining the stall force and electro-osmotic flow.

\subsection{DNA ``energy'' near the pore}

Here we calculate the quasiequilibrium energy, or the work of the stall force \cite{grosberg2010dna}, of a DNA coil with one end captured in the pore, which we use in the DNA capture model in our accompanying work \cite{PayamShuraCapture}. Attracted electrophoretically towards the pore, it is ``energetically'' more favorable for a DNA molecule to move downstream along the field lines towards the pore. Given the non-equilibrium nature of electrophoresis, this must be formulated using a quasiequilibrium energy gain, which can be defined as the work of the stall force $W$ as the DNA is brought to the pore quasistatically \cite{grosberg2010dna}. Tentatively assuming a strong suppression of electro-osmotic flow near the pore, one of us \cite{wanunu2009electrostatic} had proposed that $W\sim QV(R)$, where $Q=Nq$ was the bare charge of a coil of size $R$ and $V(R)\sim Q_{\mathrm{pore}}/R$ was the voltage at a distance $R$ from the pore. Here we show that due to the persistence of the electro-osmotic flow as described above, $W\sim \ln N$ and is in fact much smaller than $QV(R)\sim N^{1-\nu}$. 

The work of the stall force is reversible and thus path independent. We use a convenient path in which capture of one DNA end into the pore takes place in two steps. The coil is first brought to the pore in such a way that every segment is brought to an average distance $R$ from the pore. During this step, the field variations across the DNA are at most within numerical factors and thus, Eq (\ref{bucket stall force}) can be used to obtain
\begin{equation}
 W(R) \sim \int_{\infty}^{R} F_{\mathrm{st}}(r) \mathrm{d}r  \sim \eta\ \mu_E Q_{\mathrm{pore}}. \label{w(R) formula}
\end{equation} 
After the coil has arrived at the pore, one of its ends must be pulled from within the coil and brought to the pore. During this motion, the coil is pulled in such a way that any segment indexed $g$ with respect to the captured end is brought to a distance $r\sim \ell g^\nu$ from the pore. Therefore, on average, $\left(r/\ell\right)^{(1-\nu)/(\nu)} ({\mathrm{d}r }/{\ell})$ segments are brought from a distance $\sim R$ to $r$ from the pore and placed in a shell of thickness $\mathrm{d}r$. The electrophoretic pull during this step is determined by Eq (\ref{stall force per segment}), which performs a work 
\begin{equation}
 w_{\mathrm{seg}}(r)\sim\int_{R}^r f_{\mathrm{st}}(r') \mathrm{d}r' \sim \eta\ \mu_E Q_{\mathrm{pore}} \left(\frac{\ell}{r}\right)^{\frac{1}{\nu}}
\end{equation}
on a segment brought to $r$. The first segment is brought to the pore such that its near and far ends are at distances $\sim a$ and $\sim \ell$ from the pore respectively and thus, the work performed on each small piece of size $\mathrm{d} r$ of this segment is $\sim \eta \mu_E Q_{\mathrm{pore}}(\mathrm{d} r)/r$, with $a<r<\ell$. Summing over all the segments we obtain 
\begin{equation}
\begin{split}
 W_{\mathrm{cap}}\sim&\int_\ell^{R} w_{\mathrm{seg}}(r) \left(\frac{r}{\ell}\right)^\frac{1-\nu}{\nu} \frac{\mathrm{d}r }{\ell}+\int_a^\ell \eta\ \mu_E Q_{\mathrm{pore}}\frac{\mathrm{d} r}{r}  \\ \sim& \eta\ \mu_E Q_{\mathrm{pore}} \ln{\frac{R}{a}}.\label{capture energy}
 \end{split}
\end{equation}
The total energy $W=W(R)+W_{\mathrm{cap}}$ is dominated by $W_{\mathrm{cap}}$, the work performed for the DNA end to be captured while the DNA is in the vicinity of the pore, which includes a term of order $\ln N$, significant for long and flexible coils, and a term $\sim \ln{(\ell/a)}$, dominant for a rigid DNA. For a long DNA coil, the overall energy is $W\sim \eta\ \mu_E Q_{\mathrm{pore}}\ln N$, much smaller than the previously suggested value $QV(R)\sim \lambda Q_{\mathrm{pore}} N^{1-\nu}$ \cite{wanunu2009electrostatic}.

\subsection{\label{pore details}Comparison between the hydrodynamic resistance of the DNA and the pore}

We mentioned earlier that the electro-osmotic flow lines almost do not go through the pore at all even when the DNA end is captured in the pore. This is due to the high friction with the pore walls and membrane; the pore acts like a narrow pipe with a hydrodynamic resistance $\Omega_{\mathrm{p}}\sim \eta {b}/{a^4}$, and friction with the membrane results in an access resistance $\Omega_{\mathrm{a}}\sim \eta /{a^3}$ \cite{Hall01101975,access-res}. We can compare $\Omega=\Omega_{\mathrm{p}}+\Omega_{\mathrm{a}}$ to $\Omega_{\mathrm{DNA}}$, the hydrodynamic resistance of the DNA while placed at the pore, by expressing the dissipation which occurs in the DNA when a total liquid current $I=v_{\mathrm{liq}}(a) a^2$ passes through it. As we have emphasized before, the dominant portion of dissipation occurs in the vicinity of the DNA segments and is 
\begin{equation}
 \Sigma=\frac{1}{2} \left[\int_{\ell}^{R} \frac{\ell \eta\ v_{\mathrm{liq}}^2(r)}{\ln{\left(1+\frac{r_D}{d}\right)}} \frac{N(r)}{r} \mathrm{d}r+\int_a^{\ell} \frac{\eta\ v_{\mathrm{liq}}^2(r)}{\ln{\left(1+\frac{r_D}{d}\right)}} \mathrm{d}r\right], \label{DNA resistance integral}
\end{equation}
where the dissipation rate near each segment at distance $r$ is $\sim {\ell \eta\ v_{\mathrm{liq}}^2(r)}/{\ln{\left(1+{r_D}/{d}\right)}}$, and $N(r) \mathrm{d}r/r$ is the number of segments in a half spherical shell of radius $r$ and thickness $\mathrm{d}r$, with $N(r)\sim (r/l)^{1/\nu}$. The second integral accounts for the dissipation due to the first segment; every small piece of length $\mathrm{d}r$ of this segment dissipates at a rate $\sim {\eta v_{\mathrm{liq}}^2(r)\mathrm{d}r}/{\ln{\left(1+{r_D}/{d}\right)}}$. Using $\Sigma=\frac{1}{2}\ \Omega_{\mathrm{DNA}} I^2$ and $a\ll R$ we obtain
\begin{equation}
 \Omega_{\mathrm{DNA}}\sim \frac{\eta}{a^3} \frac{1}{\ln{\left(1+\frac{r_D}{d}\right)}}.
\end{equation}
For very short pores, such as those drilled on a graphene membrane, the pore resistance $\Omega_{\mathrm{p}}$ is insignificant and the access and DNA resistances are comparable. Thus, the electro-osmotic flow is maintained by circulation around the DNA as well as by sucking liquid through the pore. For long pores, typical of regular solid-state nanopores, the pore resistance is much larger than the DNA resistance and thus, the electric field drives the liquid through the coil by dominantly circulating it around the DNA and almost sucking no liquid through the pore. 

\subsection{\label{deformation}DNA conformation near the pore}

Bringing one DNA end into the pore, just like grafting a polymer from one end to a solid surface, is entropically unfavorable and causes a tension $F_{\mathrm{gr}}\sim T/r$ along the coil at a distance $r$ from the pore \cite{farkas2003dna,rowghanian2011force}. This tension holds the coil near the membrane by exerting a net force $f_{\mathrm{gr}}$ on each DNA segment, which for a segment indexed $g$ at a distance $r\sim \ell g^\nu$ from the pore is equal to 
\begin{equation}
 f_{\mathrm{gr}}(r)\sim \frac{\mathrm{d}F_{\mathrm{gr}}}{\mathrm{d} g} \sim \frac{T}{\ell}  \left(\frac{r}{\ell}\right)^{-\frac{1+\nu}{\nu}}. \label{grafting force}
\end{equation}
If we now turn on a weak electric field, an electrophoretic pull [Eq. (\ref{stall force per segment})] equal to   
\begin{equation}
 f_{\mathrm{st}}(r)\sim \frac{\eta\ \mu_E Q_{\mathrm{pore}}}{\ell}  \left(\frac{r}{\ell}\right)^{-\frac{1+\nu}{\nu}}\label{grafting force}
\end{equation}
attracts the coil towards the pore and helps hold the coil near the membrane. The stall force scales with $r$ the same way as $f_{\mathrm{gr}}$ does and as a result, it partially relaxes the grafting tension $F_{\mathrm{gr}}$. The overall tension vanishes at $\eta\ \mu_E Q_{\mathrm{pore}} \sim T$, where the entropic tension is completely relaxed by the electrophoretic pull. This occurs at a voltage 
\begin{equation}
 V_c\sim \frac{T}{\eta\ \mu_E}\frac{b}{a^2} , \label{threshold voltage} 
 \end{equation}
at which, the electric field crosses over from weak to strong. 

At $\Delta V>V_c$, the DNA gets compressed and forms concentration blobs of size $\xi_c(r)$. The gradient of the resulting non-uniform pressure $P_c(r)$ created along the coil balances the force exerted by the electric field on the coil. Assuming that the coil remains dilute enough for the liquid to flow, a spherical shell of radius $r$ and thickness $\delta r$ will be subject to an electrophoretic pull
\begin{equation}
 \delta f_{\mathrm{st}}\sim r\eta \ \mu_E E(r) \frac{\delta r}{r},
\end{equation}
where we have used the fact that the stall force is determined by the local ``bucket size'' $r$; this is likely to become less and less valid as the electric field grows and the coil becomes denser. We have included a factor $\delta r/r$ to only consider the force on the thin shell. The net force exerted on the shell due to entropic pressure is
\begin{equation}
 \delta f_{\mathrm{pr}}\sim \delta\left(P_c(r)r^2\right) \sim P_c(r) r \delta r,
\end{equation}
which holds because $\mathrm{d}P_c/\mathrm{d}r \sim P_c/r$. Using the relation between the pressure and size of the concentration blobs $P_c\sim T \xi_c^{-3}$ and balancing the two forces we obtain
\begin{equation}
 \xi_c(r)\sim \frac{T}{\eta\ \mu_E Q_{\mathrm{pore}}} r, \label{concentration blob profile}
\end{equation}
which is valid for strong fields. Since the concentration blobs cannot be smaller than the segment size $\ell$, for strong fields and close to the pore, the DNA is fully compressed with $\xi_c \sim \ell$. Eq. (\ref{concentration blob profile}) for the concentration blob size therefore applies only beyond a distance $r_c\sim \ell \eta\ \mu_E  Q_{\mathrm{pore}}/T$, which is obtained by letting $\xi_c(r_c)\sim \ell$. Thus, the electrophoretic pull per segment for strong fields close to the pore is
\begin{equation}
 f_{\mathrm{st}}(r)\sim\frac{\eta\ \mu_E Q_{\mathrm{pore}}}{\ell} \left(\frac{\ell}{r}\right)^4 \ \ , \ \ r<r_c, \label{strong force profile below r_c}
\end{equation}
which is found by dividing the stall force $F_{\mathrm{st}}(r)\sim \eta\mu_E r E(r)$ of a dense blob of size $r$ by the number of segments in that blob which is $n(r)\sim r^3/\ell^3$. At $r_c$, this crosses over to 
\begin{equation}
 f_{\mathrm{st}}(r)\sim\frac{\eta\ \mu_E Q_{\mathrm{pore}}}{\ell} \left(\frac{\ell}{r_c}\right)^4  \left(\frac{r_c}{r}\right)^{\frac{1+\nu}{\nu}} \ \ , \ \ r>r_c, \label{strong force profile below r_c}
\end{equation}
which is similarly found by dividing the stall force $F_{\mathrm{st}}(r)\sim \eta\mu_E r E(r)$ by the number $n(r)$ of the segments which are at a distance $r$ or closer from the pore. The integral which determines $n(r)$ is dominated by its upper bound if $r$ is sufficiently larger than $r_c$ and thus $n(r)$ is the volume $\sim r^3$ multiplied by the segment number density $\sim (\xi_c(r)/\ell)^{\frac{1}{\nu}}/\xi_c^3(r)$. 

Assuming the experimental conditions of the work \cite{wanunu2009electrostatic}, namely, $\lambda\sim 1\mathrm{e}/\mathrm{nm}$, $\ln{\left(1+{r_D}/{d}\right)}\sim 1$, and pore dimensions $a= 5\mathrm{nm}$ and $b=25\mathrm{nm}$, the crossover voltage is $V_c \sim 100\mathrm{mV}$. This implies that the coil deformation is negligible in those experiments and the calculated value for $W$ is valid. At the crossover between weak and strong fields, as mentioned above, the entropic tension due to holding the coil near the membrane with one end captured is fully relaxed. One way to interpret this is that all the entropic cost $\sim T\ln{N}$ of bringing the coil near the membrane is compensated by the electric field. Since at the crossover $\eta\ \mu_E Q_{\mathrm{pore}} \sim T$, we conclude that to capture the DNA end into the pore, the electrophoretic pull must perform a work $\sim \eta\ \mu_E Q_{\mathrm{pore}} \ln{N}$ on the coil, which confirms the logarithmic dependence of $W$, the DNA's quasiequilibrium energy at the pore, derived in section \ref{electrophoresis near pore}.

\section{Conclusion and Remarks}

With the aim of formulating the electrophoresis of a DNA molecule placed at the non-uniform electric field of a nanopore on a membrane, we have developed a scaling scheme which characterizes the DNA electrophoresis by describing the electro-osmotic flow through and around the DNA. This microscopic scheme is based on an analogy between electrophoresis and the ``submerged jet'' problem, where the charged liquid surrounding the DNA chain is considered as a collection of jets which cause electro-osmotic flow by receiving momentum from the electric field and injecting it into the liquid. The scheme developed here reproduces the well-known size-independent electrophoretic mobility of a DNA coil \cite{muthukumar1996theory,PhysRevLett.76.3858} and the free drain of liquid \cite{ELPS:ELPS200800257,hickey2012simulations} in a uniform field, as well as the effective charge of the DNA which scales with the size of the DNA coil and is much smaller than the DNA bare charge. Using this scheme, we then compute the electrophoretic pull (or stall force) in the non-uniform field of a pore in a membrane, and demonstrate that although the electro-osmotic flow through and around the DNA coil is somewhat reduced near the membrane, it is not completely suppressed and persists as liquid is pumped both around and through the coil. This has to be compared to the tentative picture of the previous work \cite{wanunu2009electrostatic}, in which, an almost complete suppression of the electro-osmotic flow near the pore was assumed. 

As another application of our model, we were able to calculate the liquid and electric currents driven through a polyelectrolyte gel by an external electric field and pressure gradient. Electrohydrodynamic coupling in this system was first pointed out by \citeauthor{gennes2000mechanoelectric} \cite{gennes2000mechanoelectric}, who formulated the system in terms of a linear-response theory with three independent phenomenological coefficients. This included two diagonal coefficients, namely, electric resistance, which described the electric current due to the applied voltage, and hydraulic resistance, which determined Darcy's liquid flow driven by pressure, as well as two equal off-diagonal coefficients which characterized the electric current due to pressure gradient and liquid flow driven by electric voltage. Based on our model presented here, we have developed a microscopic theory to determine these coefficients. The hallmark of our results, in contrast to the previous work \cite{doi:10.1021/ma047944j}, is that we were able to explore the dependence on the screening radius, or more generally, on the thickness $r_D$ of the charged sleeve surrounding the chains, thus relaxing the unphysical assumption implicit in the previous work that $r_D$ is much larger than the network mesh size. 

\begin{acknowledgments}
This research was supported in part by the National Science Foundation under Grant No. NSF PHY11-25915. We would like to thank the Kavli Institute for Theoretical Physics in Santa Barbara where part of this work was done. The work of A. Y. G. was supported in part by a grant from the U.S.-Israel Binational Science Foundation, and P. R. was supported by the National Science Foundation under Grant No. NSF PHY-0424082. We would like to also acknowledge constructive discussions with Y. Rabin, B. Shklovskii, and L. Lizana.
\end{acknowledgments}


\end{document}